\documentclass[pre, showpacs, showkeys, nofootinbib]{revtex4}
\usepackage{latexsym}
\usepackage{amsmath}
\usepackage[dvips]{graphicx}

\begin {document}
\title{Stochastic properties of systems controlled
by autocatalytic reactions I}
\author{L. \surname{P\'al}}
\email[Electronic address:]{lpal@rmki.kfki.hu} \affiliation{KFKI
tomic Energy Research Institute H-1525 Budapest 114, POB 49
Hungary}

\begin {abstract}
We analyzed the stochastic behavior of systems controlled by
autocatalytic reaction $A + X \rightarrow X +X$. Assuming the
distribution of reacting particles in the system volume to be
uniform, we introduced the notion of the point model of reaction
kinetics, and derived a system of differential equations for
probabilities of finding $n=0,1,\ldots$ autocatalytic particles at
a given time moment. It has been found that the kinetic law of the
mass action cannot be supported by stochastic model.
\end{abstract}



\maketitle

\section*{Introduction}

The system denoted by $\mathcal S$ is defined as an aggregation of
particles capable for autocatalytic reactions. Symbols $X$ and $A$
are used for notations of \textit{autocatalytic and substrate
particles}, respectively. The system is "open" for particles $A$,
when it is in contact with a large reservoir which keeps constant
the number of particles $A$. In contrary, the system is "closed",
when particles cannot be injected in or extracted from the system.
We suppose that the system is always closed for the autocatalytic
particles $X$.

As known, there are many papers and
books~\cite{rice85,kotomin96,gardiner83,kampen04} dealing with the
influence of spacial (mostly diffusive) motion of particles on the
kinetics of reactions, however, in many cases the complexity of
the problem did not allow to obtain exact results. In order to
analyze one of the decisive factors determining the stochastic
nature of autocatalytic reactions, in the present paper we do not
deal with the diffusion of particles, instead we introduce the
notion of \textit{the point model of reaction kinetics}, similarly
to that used in the theory of neutron chain reactors. It is to
note that the point model of the reaction kinetics is based on a
sever assumption according to that the reacting particles are
distributed uniformly in the whole volume of the system. In this
case, it can be stated that the probability of a reaction between
two particles is proportional to the product of their actual
numbers in the system. This approach is a highly simplified but
useful model for systems, in which the number of particles is
small, and the largest distant within the system is smaller than
the characteristic length of interaction between the particles.

I the sequel we will study closed systems controlled by random
process in which the {\em particles $X$ catalyze the conversion of
particles $A$ into further $X$'s}. Changing adiabatically the
number of particles $A$  with help of a suitable reservoir, we can
investigate the influence of the environment on the system
behavior.

We say that a system is in a \textit{living state} when the
conditions for autocatalytic reactions are existing, while a
system is in a \textit{dead state} when there is no possibility
for autocatalytic reactions. The random time spent by a system in
the living state is \textit{the lifetime of the system}. One of
the aims of this paper is the determination of the probability
distribution of the lifetime and the study of its random
properties.

The organization of the paper as follows. To elucidate the
approach that we will follow in the present work, in Section 1 we
deal with systems controlled by the autocatalytic reaction $A + X
\rightarrow X +X$. We derive and solve a system of differential
equations determining the  probabilities to find $n=0, 1, \ldots,
N_{A}$ new $X$ particles arisen from $A$ particles during the time
interval $(0,t)$. Since the system is closed, the number of
particles $A$ is decreasing with time, and finally becomes zero,
i.e. the system reaches its dead state. By using the generating
function equations, in Section 2 we show that the stochastic
approach brings about an equation for the mean value of the number
of autocatalytic particles completely different from the classical
rate equation based on the kinetic law of the mass action. In
order to obtain some kind of solution of the hierarchical equation
system derived for moments, \textit{the moment-closure
(decoupling) approximation} has been applied by  many
authors~\cite{whittle57,nishiyama74}, however, the consequences of
the this procedure were hardly investigated. Since we succeeded to
obtain exact results for the time dependence of moments, in
Section 3 we analyzed the error caused by the decoupling
approximation. Finally, in Section 4  we study the random
properties of the system lifetime.

\section{Systems controlled by reaction $A + X \rightarrow 2 X$}

We would like to study the properties of \textit{closed systems}
${\mathcal S}$ of constant volume $V$ controlled by the reaction
\[ A +  X \stackrel{\mathrm{k_{c}}}{\rightarrow}  X +  X. \] Denote by
$N_{A}$ and $N_{X}$ the numbers of particles $A$ and $X$,
respectively at time instant $t=0$. As the particles $X$ convert
$A$ into further $X$'s, so the number of particles $A$ strictly
decreases, and finally, the autocatalytic process is stopped.
Obviously, in this case the system contains only $X$ particles,
and this is the state which is called \textit{dead state}.

\subsection{Kinetic rate equation}

Denote by $m(t)$ the expectation value of the number of particles
$A$ converted into $X$ during the time interval $[0, t]$.
Obviously,
\begin{equation} \label{x1}
m_{A}(t) = N_{A} - m(t), \;\;\;\;\;\; \text{and} \;\;\;\;\;\;
m_{X}(t) = N_{X} + m(t)
\end{equation}
are the average number of particles $A$ and $X$, respectively at
the time moment $t \geq 0$. Introducing the notations
\[  a(t) = \frac{m(t)}{V}, \;\;\;\;\;\; c_{A}(0) =
\frac{N_{A}}{V}, \;\;\;\;\; \mbox{and} \;\;\;\;\;\; c_{X}(0)
=\frac{N_{X}}{V}, \] according to the kinetic law of mass action
we can write the rate equation in the following form:
\[ \frac{da(t)}{dt} = k_{c} [c_{A}(0) - a(t)]
\;[c_{X}(0) + a(t)]. \]  Taking into account the initial condition
$a(0) = 0$, and using the abbreviation $u = k_{c}\;t/V$, one
obtains that
\begin{equation} \label{x2}
a(u) = c_{A}(0) \frac{1 - e^{-N_{0}u}}{1 + \rho\;e^{-N_{0}u}},
\end{equation}
where
\[ \rho = \frac{N_{A}}{N_{ X}}\;\;\;\;\;\; \text{and}
\;\;\;\;\; N_{0} = N_{A}+ N_{X}. \] For the sake of comparison
with the results of calculations of the stochastic model, let us
introduce the ratios:
\begin{equation} \label{x3}
r_{A}(u) = \frac{m_{A}(u)}{N_{A}} = (1 +
\rho)\;\frac{e^{-N_{0}u}}{1 + \rho\;e^{-N_{0}u}},
\end{equation}
and
\begin{equation} \label{x4}
r_{X}(u) = \frac{m_{X}(u)}{N_{X}} = (1 + \rho)\;\frac{1}{1 +
\rho\;e^{-N_{0}u}}.
\end{equation}

\subsection{Stochastic model}

Denote the number of particles $A$ converted into particles $X$
during the time interval $[0, t]$ by the random function $\xi(t)
\in {\mathcal Z_{+}}$, where ${\mathcal Z_{+}}$ is the set of
nonnegative integers. At the same time, note that $\xi(t)$ is the
number of new $X$ particles appearing in $[0, t]$. Let
\begin{equation} \label{x5}
{\mathcal P}\{ \xi(t)=n \vert \xi(t)=0\} = p(t,n), \;\;\;\;\;\; 0
\leq n \leq N_{A}
\end{equation}
be the probability to find $n$ new $X$ particles at the time
moment $t \geq 0$ in the system ${\mathcal S}$ provided that at
$t=0$ the number of new particles $X$ was zero and the system
contained $N_{A}$ particles of type $A$ and $N_{X}$ particles of
type $X$. The pair of integers $\{N_{A},\; N_{X}\}$ denotes the
\textit{initial state of the system $\mathcal S$}. Assume that
$\alpha\;\Delta t + o(\Delta t)$ is the probability that any of
particles $X$ brings about a reaction with any of particles $A$.
If the numbers of $A$ and $X$ particles in the system at time
instant $t \geq 0$ are $N_{A} - n$ and $N_{X} + n$, respectively,
then
\begin{equation} \label{x6}
h_{n} \Delta t + o(\Delta t) = \alpha\;[N_{{A}} - n]\; [N_{{X}} +
n] \Delta t + o(\Delta t)
\end{equation}
is the probability that the reaction $ A +  X
\stackrel{\mathrm{k_{c}}}{\longrightarrow} 2\;X$ is realized in
the time interval $(t, t + \Delta t)$. It is to note that $\alpha
= k_{c}/V$ and $h_{N_{A}} = 0$. By using the expression (\ref{x6})
one can write
\[ p(t+\Delta t, n) = p(t, n)(1 - h_{n} \Delta t) + h_{n-1} p(t,
n-1) \Delta t + o(\Delta t), \] and obtain immediately the
equation
\begin{equation} \label{x7}
\frac{dp(t, n)}{dt} = - h_{n} p(t, n) + h_{n-1} p(t, n-1),
\;\;\;\;\;\; 1 \leq n \leq N_{A}.
\end{equation}
If $n=0$, then
\begin{equation} \label{x8}
\frac{dp(t, 0)}{dt} = - h_{0} p(t, 0),
\end{equation}
and taking into account the initial condition $p(0, 0) = 1$ one
has
\begin{equation} \label{b9}
p(t, 0) = e^{-h_{0}t} = e^{-N_{A}N_{X}\;u}.
\end{equation}
By introducing the Laplace transforms
\[ U_{n}(s) = \int_{0}^{\infty} e^{-st}\;p(t, n)\;dt, \;\;\;\;\;\;
n=1,2, \ldots, N_{A}, \] since $p(0, n) = 0$, if $n > 0$, we
obtain from Eq. (\ref{x7}) the recursive relation
\[ (s + h_{n})\;U_{n} = h_{n-1}\;U_{n-1}(s), \;\;\;\;\;\;
n > 0, \] which can be easily solved by starting with
\begin{equation} \label{x10}
U_{0}(s) = \frac{1}{s+h_{0}}.
\end{equation}
The solution can be written into the form:
\begin{equation} \label{x11}
U_{n}(s) = \frac{h_{0}h_{1} \cdots h_{n-1}}{(s+h_{0})(s+h_{1})
\cdots (s+h_{n})}, \;\;\;\;\;\; n=1,2, \ldots, N_{A}.
\end{equation}
In order to obtain the probabilities $p(t, n), \;\; 1 \leq n \leq
N_{A}$, the next step seems to be very simple: one has to expand
$U_{n}(s)$ into the series of partial fractions, and then to
perform the inverse Laplace transformation. However, the task is
slightly more complex.

If $N_{A} - N_{X} = n_{c} \leq 0$, then it is obvious that the
quantities $h_{0}, h_{1}, \ldots, h_{n}$ are all different and so,
we can write
\begin{equation} \label{x12}
U_{n}(s) = \sum_{k=0}^{n} \frac{C_{nk}}{s+h_{k}},
\end{equation}
where
\begin{equation} \label{x13}
C_{nk} = \frac{h_{0}h_{1} \cdots
h_{n-1}}{(h_{0}-h_{k})(h_{1}-h_{k}) \cdots
(h_{k-1}-h_{k})(h_{k+1}-h_{k}) \cdots (h_{n}-h_{k})},
\end{equation}
\[ n = 1, 2, \ldots, N_{A} \;\;\;\;\;\; \mbox{and}
\;\;\;\;\;\; k = 0, 1, \ldots, n, \] and since $C_{00} = 1$, we
obtain
\begin{equation} \label{x14}
p(t, n) = \sum_{k=0}^{n} C_{nk}\;e^{-h_{n}t}, \;\;\;\;\;\; n=0,1,
\ldots, N_{A}.
\end{equation}

If $N_{A} - N_{X} = n_{c} > 0$, then it can be easily proved the
following statement: if $n_{c}$ is even, i.e. $n_{c}=2k_{c}$, then
\[ h_{n_{c}}=h_{0},\; h_{n_{c}-1}=h_{1}, \ldots, h_{k_{c}+1} =
h_{k_{c}-1}, \] while if $n_{c}$ is odd, i.e. $n_{c}=2k_{c}+1$,
then
\[ h_{n_{c}}=h_{0}, \; h_{n_{c}-1}=h_{1}, \ldots, h_{k_{c}+1} =
h_{k_{c}}. \] Let us denote the denominator of $U_{n}(s)$ by
\[ V_{n}(s) = (s+h_{0})(s+h_{1}) \cdots (s+h_{n}), \]
and introduce the notation
\[ Q(j, s) = \prod_{i=0}^{j}(s+h_{i})^{2}. \]
In the case of $n > n_{c}$ we can show that
\[ V_{n}(s) = \left\{\begin{array}{ll}
Q(k_{c}-1, s)(s+h_{k_{c}})(s+h_{n_{c}+1})
\cdots (s+h_{n}), & \mbox{if $n_{c}=2k_{c}$,} \\
\mbox{} & \mbox{} \\
Q(k_{c}, s)(s+h_{n_{c}+1}) \cdots (s+h_{n}), & \mbox{if
$n_{c}=2k_{c}+1$,}
\end{array} \right. \]
while in the case of $n=n_{c}$ we obtain
\[ V_{n_{c}}(s) = \left\{\begin{array}{ll}
Q(k_{c}-1, s)(s+h_{k_{c}}), & \mbox{if $n_{c}=2k_{c}$,} \\
\mbox{} & \mbox{} \\
Q(k_{c}, s), & \mbox{if $n_{c}=2k_{c}+1$.}
\end{array} \right. \]
\begin{figure} [ht!]
\protect \centering{
\includegraphics[height=10cm, width=7cm]{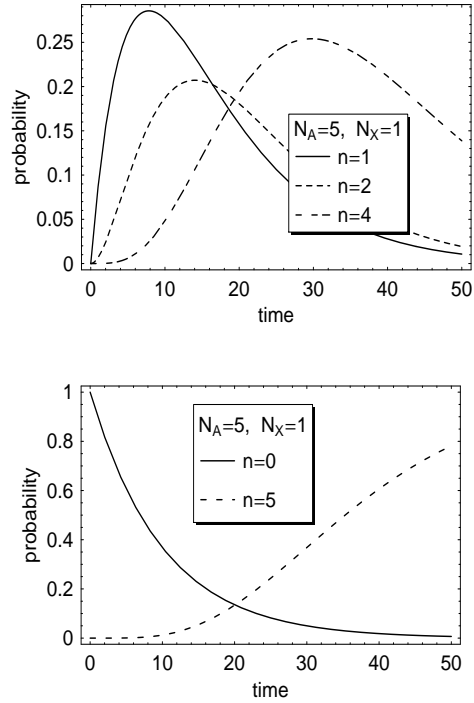}}\protect
\vskip 0.2cm \protect \caption{\label{fig1} {\footnotesize
Probabilities of finding $n=0,1,2,4,5$ new $X$ particles at time
moment $t$ provided that the initial state of the system was
\{5,1\}.}}
\end{figure}
When $n < n_{c}$ and $0 < n \leq k_{c}$, then we have
\[ V_{n}(s) = \prod_{i=0}^{n} (s+h_{i}) \]
for both, even and odd values of $n_{c}$, but in the case of
$k_{c} < n < n_{c}$ we should use the formula:
\[ V_{n}(s) = \left\{\begin{array}{ll}
\prod_{i=0}^{n_{c}-n-1}(s+h_{i})\;Q(n-k_{c}-1, s)(s+h_{k_{c}}),
& \mbox{if $n_{c}=2k_{c}$,} \\
\mbox{} & \mbox{} \\
\prod_{i=0}^{n_{c}-n-1}(s+h_{i})\;Q(n-k_{c}-1, s), & \mbox{if
$n_{c}=2k_{c}+1$.}
\end{array} \right. \]

For the sake of illustration it seems to be useful to write down
the equations of $U_{n}(s)$ when $N_{A}=5$ and $N_{X}=1$. One
obtains
\begin{eqnarray}
U_{5}(s) & = & \frac{h_{0}^{2} h_{1}^{2}
h_{2}}{(s+h_{0})^{2}(s+h_{1})^{2}(s+h_{2})s}, \;\;\;\;\;\;
U_{0}(s) = \frac{1}{s+h_{0}} \nonumber \\
U_{4}(s) & = & \frac{h_{0} h_{1}^{2}
h_{2}}{(s+h_{0})^{2}(s+h_{1})^{2}(s+h_{2})}, \;\;\;\;\;\;\;\;
U_{1}(s) = \frac{h_{0}}{(s+h_{0}) (s+h_{1})}, \nonumber \\
U_{3}(s) & = & \frac{h_{0} h_{1}
h_{2}}{(s+h_{0})(s+h_{1})^{2}(s+h_{2})}, \;\;\;\;\;\;\;\;\;
U_{2}(s) = \frac{h_{0} h_{1}}{(s+h_{0})(s+h_{1})(s+h_{2})}.
\nonumber
\end{eqnarray}
One can see in FIG.~\ref{fig1} that the probability $p(t,n)$
versus $t$ curves with exception of $n=0$ and $n=5$ have a
well-defined maximum the location of which is increasing with $t$.

\section{Generating function equation}

In order to make easier the derivation of equations for  moments
of the random function $\xi(t)$ it is worthwhile to introduce the
generating function
\begin{equation} \label{x15}
\tilde{g}(s, z) = \sum_{n=0}^{N_{A}} U_{n}(s)\;z^{n}.
\end{equation}
It can be proved that $\tilde{g}(s, z)$ satisfies the following
equation:
\begin{equation} \label{x16}
\alpha z^{2}(1-z)\;\frac{d^{2}\tilde{g}(s, z)}{dz^{2}} - \alpha
(n_{c}-1) z (1-z)\;\frac{d\tilde{g}(s, z)}{dz} - [s +
h_{0}(1-z)]\;\tilde{g}(s, z) + 1 = 0.
\end{equation}
In many cases the use of the exponential generating function
\begin{equation} \label{x17}
\tilde{g}_{exp}(s, y) = \sum_{n=0}^{N_{A}} U_{n}(s)\;e^{n y}
\end{equation}
is more advantageous, therefore we write it down also:
\begin{equation} \label{x18}
\alpha (e^{y}-1)\;\frac{d^{2}\tilde{g}_{exp}(s, y)}{dy^{2}} -
\alpha n_{c} (e^{y}-1)\; \frac{d\tilde{g}_{exp}(s, y)}{dy} + [s -
h_{0}(e^{y}-1)]\;\tilde{g}_{exp}(s, y) - 1 = 0.
\end{equation}
We can see that
\[ \lim_{z \rightarrow 1} \tilde{g}(s, z) = \lim_{y \rightarrow 0}
\tilde{g}_{exp}(s, y) = \frac{1}{s}, \] and it is elementary to
show that the generating function
\[ g(t, z) = \hat{{\mathcal L}}^{-1}\{\tilde{g}(s, z)\}, \]
where $\hat{{\mathcal L}}^{-1}$ is the operator of the inverse
Laplace transformation, satisfies the equation
\begin{equation} \label{x19}
\frac{\partial g(t, z)}{\partial t} = - h_{0}(1-z)\;g(t, z) -
\alpha (n_{c}-1) z (1-z)\;\frac{\partial g(t, z)}{\partial z} +
\alpha z^{2}(1-z)\;\frac{\partial^{2}g(t, z)}{\partial z^{2}},
\end{equation}
with \[ g(0, z) = 1, \;\;\;\;\;\; \mbox{and} \;\;\;\;\;\; g(t, 1)
= 1.\] Similarly, the exponential generating function
\[ g_{exp}(t, y) = \hat{{\mathcal L}}^{-1}\{\tilde{g}_{exp}(s, z)\} \]
is nothing else then the solution of the equation
\begin{equation} \label{x20}
\frac{\partial g_{exp}(t, y)}{\partial t} = - h_{0}(1-e^{y})\;
g_{exp}(t, y) - \alpha n_{c}(1-e^{y})\;\frac{\partial g_{exp}(t,
y)}{\partial y} + \alpha (1-e^{y})\;\frac{\partial^{2}g_{exp}(t,
y)}{\partial y^{2}},
\end{equation}
with \[ g_{exp}(0, y) = 1, \;\;\;\;\;\; \mbox{and} \;\;\;\;\;\;
g_{exp}(t, 0) = 1.\]

\subsection{Expectation value}

The numbers of particles $X$ and $A$ at time instant $t \geq 0$
are given by
\[ \xi_{X}(t) = N_{X} + \xi(t), \;\;\;\;\;\;
\text{and} \;\;\;\;\;\;  \xi_{A}(t) = N_{A} - \xi(t),\]
respectively. From these one obtains
\begin{eqnarray} \label{x21}
{\bf E}\{\xi_{X}(t)\} = m_{X}(t) & = & N_{X} + {\bf E}\{\xi(t)\}
\\ \label{x22} {\bf E}\{\xi_{A}(t)\} = m_{A}(t)
& = & N_{A} - {\bf E}\{\xi(t)\},
\end{eqnarray}
where
\begin{equation} \label{x23}
{\bf E}\{\xi(t)\} = m_{1}(t) = \left[\frac{\partial g(t,
z)}{\partial z}\right]_{z=1} = \left[\frac{\partial g_{exp}(t,
y)}{\partial y}\right]_{y=0}.
\end{equation}
By using the equation (\ref{x20}) after elementary algebra we
obtain
\begin{equation} \label{x24}
\frac{dm_{1}(t)}{dt} =  h_{0} + \alpha n_{c} m_{1}(t) - \alpha
m_{2}(t),
\end{equation}
where
\[ m_{2}(t) = {\bf E}\{\xi^{2}(t)\}, \]
and this equation can be rewritten in the form:
\begin{equation} \label{x25}
\frac{dm_{1}(t)}{dt} = \alpha[N_{A}-m_{1}(t)]\; [N_{X} + m_{1}(t)]
- \alpha {\bf D}^{2}\{\xi(t)\}.
\end{equation}
As seen the appearance of the variance ${\bf D}^{2}\{\xi(t)\}$
brings about the loss of validity of the \textit{kinetic law of
the mass action}. If ${\bf D}^{2}\{\xi(t)\} \approx 0$, i.e. if
$m_{2}(t) \approx [m_{1}(t)]^{2}$, then the deterministic kinetic
equation can be applied.

\begin{figure} [ht!]
\protect \centering{
\includegraphics[height=8cm, width=12cm]{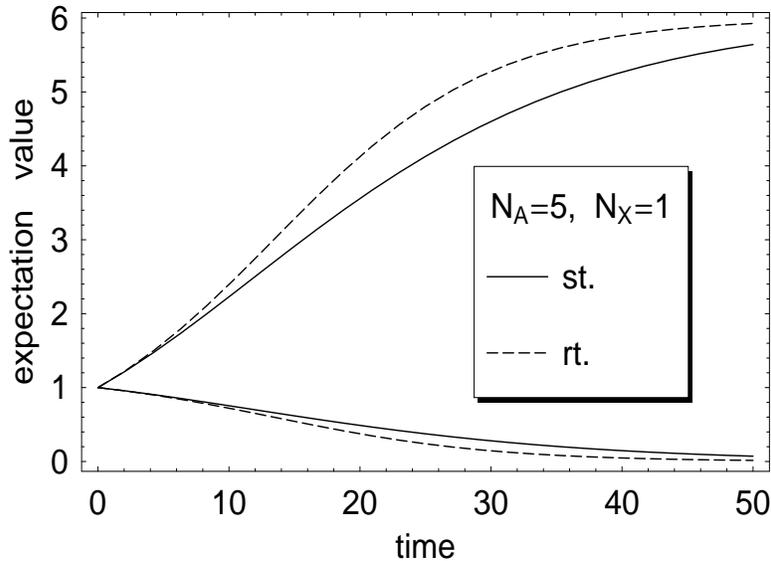}}\protect
\vskip 0.2cm \protect \caption{\label{fig2}{\footnotesize
Expectation values of the relative numbers of particles of type
$X$ (upper curves) and of type $A$ (lower curves) versus time
provided that the initial state of the system was $\{5,1\}$.}}
\end{figure}

The equation (\ref{x24}) shows the hierarchical structure of
momentum equations. If we would like to calculate $m_{1}(t)$ by
solving the equation (\ref{x24}), we see that for this we need the
second moment $m_{2}(t)$. The equation of the second moment
contains the third moment etc. In this way we can obtain the time
dependence of the average number of ${X}$ particles $m_{1}(t)$
when we solve the hierarchical system of equations which seems to
be rather difficult. Therefore, in the practice the \textit{method
of decoupling}~\footnote{The decoupling is the substitution of the
moment $m_{k}(t)$ with an expression containing moments with
indices smaller than $k$.} has been often used for finding an
"approximate" solution, however the consequences of this procedure
were hardly investigated. In the next Section we try to give some
elementary analysis of the problem.

In the present case it is fortunate that we can determine the
probabilities $p(t, n), \;\; n=0,1,\ldots, N_{A}$ from
(\ref{x11}), and so we can calculate directly {\em the exact time
dependence of moments}
\begin{equation} \label{x26}
m_{k}(t) = {\bf E}\{[\xi(t)]^{k}\} = \sum_{n=0}^{N_{A}}
n^{k}\;p(t, n).
\end{equation}
The expectation values of the relative numbers of $X$ and $A$
particles, i.e. the mean values of the random functions
\[ \rho_{X}(t) = 1 + \frac{\xi(t)}{N_{X}},
\;\;\;\;\;\; \text{and} \;\;\;\;\;\; \rho_{A}(t) = 1 -
\frac{\xi(t)}{N_{A}} \] versus time can be seen in
FIG.~\ref{fig2}. The upper curves refer to the ${X}$, while the
lower curves to the ${A}$ particles. The continuous and dashed
curves show the results of calculations for the stochastic and
deterministic models, respectively. As expected the difference
between the stochastic and deterministic description is relatively
small at the beginning and at the end of the process.

\subsection{Variance}

We have seen that the occurrence of the variance ${\bf
D}^{2}\{\xi(t)\}$ in the differential equation of (\ref{x25}) is
the source of the invalidity of the kinetic law of the mass
action. Therefore, it seems to be worthwhile to look at the time
dependence of the variance and relative variance of the number of
$X$ particles.

\begin{figure} [ht!]
\protect \centering{
\includegraphics[height=12cm, width=8cm]{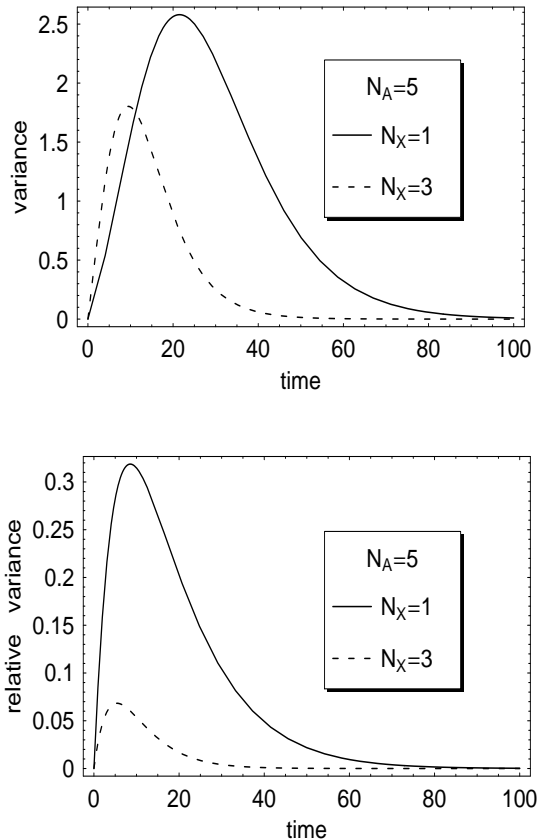}}\protect
\vskip 0.2cm \protect \caption{\label{fig3}{\footnotesize Time
dependence of the variance ${\bf D}^{2}\{\xi_{X}(t)\}$ and the
relative variance ${\bf D}^{2}\{\xi_{X}(t)\}/{\bf
E}^{2}\{\xi_{X}(t)\}$ in two initial states: ${\mathcal S}_{0} =
\{5,1\}, \; \{5,3\}$.}}
\end{figure}

Since $\xi_{X}(t) = N_{X} + \xi(t)$ and $\xi_{A}(t) = N_{A} -
\xi(t)$, it is trivial that
\[ {\bf D}^{2}\{\xi_{X}(t)\} = {\bf D}^{2}\{\xi_{A}(t)\} =
{\bf D}^{2}\{\xi(t)\}, \] where
\[ {\bf D}^{2}\{\xi(t)\} = \sum_{n=0}^{N_{A}} \left[n -
m_{1}(t)\right]^{2}\;p(t, n). \] By using the inverse Laplace
transform of the expression (\ref{x11}), we can calculate the
variance ${\bf D}^{2}\{\xi(t)\}$. The results of calculations are
plotted in FIG.~\ref{fig3}. Observe, the increase of the number of
starting ${X}$ particles from $1$ to $3$ brings about a
significant decrease of the relative variance. The fluctuation of
$\xi(t)$  increases sharply at the beginning of the process, and
decreases slowly with increasing time. After elapsing a
sufficiently large time we see that the fluctuation is negligible,
and it means that the kinetic rate equation can be regarded in
long time limit as an acceptable approximation.

\section{Moment-closure approximation}

In order to show the consequences of heuristic methods applied
often for obtaining an "approximate" solution of the hierarchical
equation systems, let us derive the exact equations for $m_{1}(t)$
and $m_{2}(t)$. It follows from (\ref{x20}) that
\begin{equation} \label{a1}
\frac{dm_{1}(t)}{dt} = h_{0} + \alpha n_{c}\;m_{1}(t) -
\alpha\;m_{2}(t)
\end{equation}
and
\begin{equation} \label{a2}
\frac{dm_{2}(t)}{dt} = h_{0} + (2h_{0}+\alpha n_{c})\;m_{1}(t) -
\alpha (2n_{c}-1)\;m_{2}(t) - 2\alpha\;m_{3}(t).
\end{equation}
We see that further equations are needed for $m_{3}(t), m_{4}(t),
\ldots,$ and this makes the problem rather complex, however, the
stationary values $m_{k}^{(st)}=\lim_{t \rightarrow \infty}
m_{k}(t), \; k=1,2, \ldots$ can be easily obtained.  Since
$m_{1}^{(st)}=N_{A}$, from (\ref{a1}) it follows that
$m_{2}^{(st)}=N_{A}^{2}$, and by induction it can be proved that
\[ m_{k}^{(st)} = N_{A}^{k}, \;\;\;\;\;\; k=1,2, \ldots
\;\; .\] If we would like to determine the time dependence of the
mean value $m_{1}(t)$, then we should solve the hierarchial system
of moment equations. The usual treatment of this problem is to use
the moment-closure approximation, i.e. to truncate the
hierarchical  equation system. However, this procedure may bring
about hardly foreseeable consequences.

For example, if we substitute $m_{3}(t)$ in Eq. (\ref{a2}) with
the product $m_{1}(t)\;m_{2}(t)$, then we obtain two differential
equations for $m_{1}(t)$ and $m_{2}(t)$, namely~\footnote{Though
these equations are different from the exact equations (\ref{a1})
and (\ref{a2}), for the sake of simplicity, we are using the same
notations for the approximate first and second moments.}
\[ \frac{dm_{1}(t)}{dt} = h_{0} + \alpha n_{c}\;m_{1}(t) -
\alpha\;m_{2}(t), \] and \[ \frac{dm_{2}(t)}{dt} = h_{0} +
(2h_{0}+\alpha n_{c})\;m_{1}(t) +\alpha (2n_{c}-1)\;m_{2}(t) -
2\alpha\;m_{1}(t)\;m_{2}(t). \] Eliminating $m_{2}(t)$ in the
second equation we can write that
\begin{equation} \label{a3}
\frac{d^{2}m_{1}(u)}{du^{2}} + 2n_{c}[N_{A}-m_{1}(u)]\;
[N_{X}+m_{1}(u)] + (1-3n_{c})\frac{dm_{1}(u)}{du} + 2m_{1}(u)
\frac{dm_{1}(u)}{du} = 0,
\end{equation}
where $u=\alpha t$. We have to solve this equation, by taking into
account the initial conditions
\begin{equation} \label{a4}
m_{1}(0) = 0, \;\;\;\;\;\; \mbox{and} \;\;\;\;\;\;
\left[\frac{dm_{1}(u)}{du}\right]_{u=0} = N_{A}\;N_{X}.
\end{equation}
For the sake of simplicity we assume that $N_{A} = N_{X} = M$,
i.e. $n_{c}=0$.  In this case we obtain
\[ \frac{d}{du}\left(\frac{dm_{1}}{du} + m_{1} + m_{1}^{2}\right)
= 0. \] It follows from this
\[ \frac{dm_{1}}{du} + m_{1} + m_{1}^{2} = C, \]
where $C$ can be determined by applying the conditions (\ref{a4}).
As seen, $C = M^{2}$, and so we have
\[ \frac{dm_{1}}{du} = - ( m_{1}^{2} + m_{1} - M^{2}), \]
the solution of which is given by
\begin{equation} \label{a5}
m_{1}(u) = 2 M^{2} \frac{1 - e^{-\lambda_{M} u}}{\lambda_{M} (1 +
e^{-\lambda_{M} u}) + 1 - e^{-\lambda_{M} u}},
\end{equation}
where $\lambda_{M} = \sqrt{4M^{2} + 1}$. The first thing that can
be seen immediately is
\[ \lim_{u \rightarrow \infty} m_{1}(u) = M\;\frac{1}{\sqrt{1 +
1/(2M)^{2}} + 1/2M}, \] and this limit  value does not correspond
to the exact relation $\lim_{u \rightarrow \infty} m_{1}(u) = M$.

\begin{figure} [ht!]
\protect \centering{
\includegraphics[height=12cm, width=8cm]{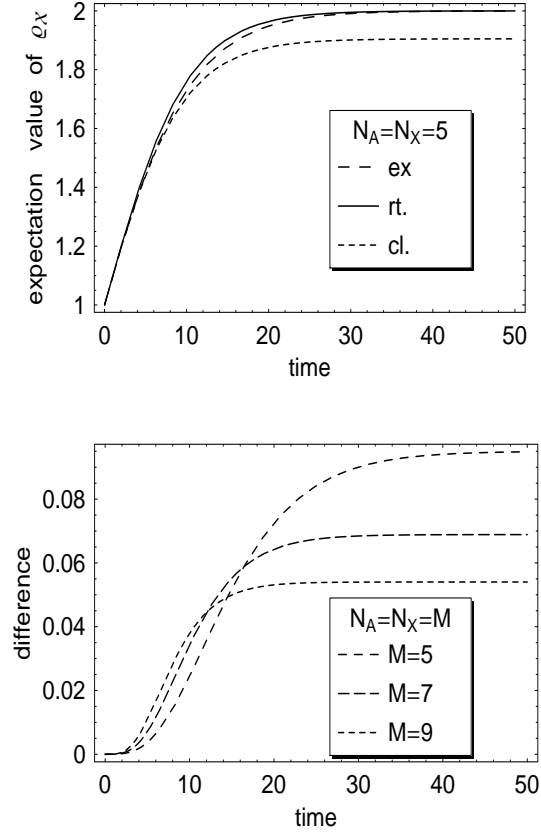}}\protect
\vskip 0.2cm \protect \caption{\label{fig4}{\footnotesize In the
upper part the time dependence of $\mathbf{E}\{\rho_{X}(u)\}$
calculated by exact "ex" and approximate "cl" methods is shown.
For the sake of comparison the curve "rt" obtained from
Eq.(\ref{x4}) is also plotted . In the lower part the difference
between the curves "ex" and "cl" vs. $u=\alpha t$ is shown for
three values of $M$.}}
\end{figure}

In the upper part of FIG.~\ref{fig4} we plotted the exact "ex" and
approximate "cl" values of $\mathbf{E}\{\rho_{X}(u)\}$) versus
$u=\alpha t$, and for the sake of comparison the curve "rt"
obtained from Eq.(\ref{x4}) is also plotted. It is remarkable that
the decoupling $m_{3}(t) \approx m_{1}(t) m_{2}(t)$ brings about
larger error than the rate equation (\ref{x4}) which corresponds
to the decoupling $m_{2}(t) \approx m_{1}(t) m_{1}(t)$. As seen,
the difference between the curves "ex" and "cl"  vs. $u$ is
negligible at the beginning and at the end of the process. The
lower part of FIG.~\ref{fig4} shows the time dependence of the
differences between the curves "ex" and "cl" for three values of
$M$.  Note, the difference decreases sharply with increasing $M$.
Summarizing, we can state that the decoupling $m_{3}(t) \approx
m_{1}(t)\;m_{2}(t)$ brings about an unacceptable error in the
calculation of  $m_{1}(t)$.

However, there is an other way to truncate the hierarchical
equation system. We can calculate the cumulants of $\xi(t)$ from
the derivatives of $K(t, y) = \log \;g_{exp}(t, y)$ at $y=0$. As
known, the cumulants
\[ \kappa_{k}(t) = \left[\frac{\partial^{k} K(t, y)}
{\partial y^{k}}\right]_{y=0} \] are zero in the case of normal
distribution if $k \geq 3$. Although the random function $\xi(t)$
does not follow the normal distribution at any $t \geq 0$, in
contrary to this we assume that $\kappa_{3}(t)=0$, and hence we
use the following decoupling:
\begin{equation} \label{a6}
m_{3}(t) \approx 3 m_{2}(t) m_{1}(t) - 2 m_{1}^{3}(t).
\end{equation}

\begin{figure} [ht!]
\protect \centering{
\includegraphics[height=7cm, width=10.5cm]{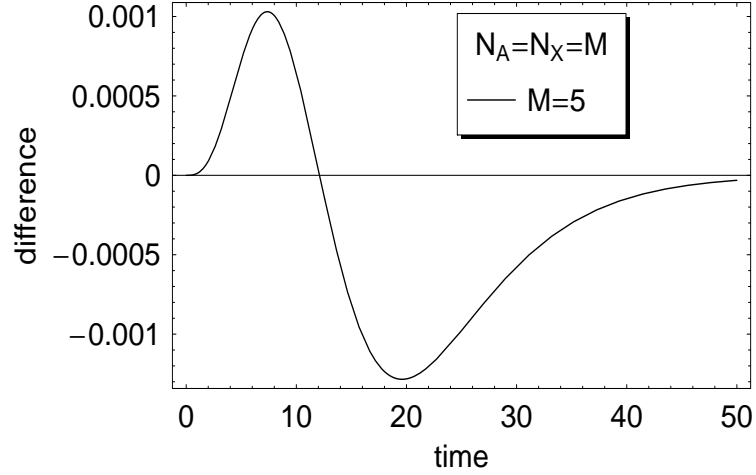}}\protect
\vskip 0.2cm \protect \caption{\label{fig5}{\footnotesize
Dependence of the difference between the exact and approximate
solutions on $u=\alpha t$. The approximate solution corresponds to
the closure $\kappa_{3}(t)=0$. }}
\end{figure}
This closure is called by some authors \cite{nishiyama74} "normal
approximation". In order to simplify the calculations, we are
dealing with the case when $N_{A} = N_{X} = M$. By using the
expression (\ref{a6}), we obtain from (\ref{a1}) and (\ref{a2})
the equations
\[ \frac{1}{\alpha}\;\frac{dm_{1}(t)}{dt} = M^{2} - m_{2}(t)\]
and
\[ \frac{1}{\alpha}\;\frac{dm_{2}(t)}{dt} = M^{2} - m_{2}(t)\;[1 + 6m_{1}(t)] +
2m_{1}(t)[M^{2} + 2m_{1}^{2}(t)], \]

As expected, one can easily show that the stationary stable
solution is $m_{1}^{(st)} = M$. The time dependence has been
calculated numerically. FIG.~\ref{fig5} shows the dependence of
the difference between the exact and the "normal" approximate
solutions  on $u=\alpha t$. One can observe that the difference is
surprisingly small even if few particles ($M=5$) are participating
in the process, however one cannot conclude that the closure
$\kappa_{3}(t)=0$ is always the best.

\section{Lifetime of the system}

Assume that the \textit{system ${\mathcal S}$ is "live"} when it
contains at least $1$ particle of type $X$ and the number of
particles $A$ is larger than zero. In the course of the process
the $X$ particles are converting the $A$ particles into $X$
particles until $A$ particles can be found in the system. Denote
by $\theta$ that random moment when the last $A$ particle is
converted into $X$. The random variable $\theta$ is called
\textit{lifetime of the system} ${\mathcal S}$. It is obvious that
the event $\{\theta \leq t\}$ is equivalent to the event
$\{\xi(t)=N_{A} \vert {\mathcal S}_{0}\}$, where $N_{A}$ and
$N_{X}$ are larger than zero. Consequently
\begin{equation} \label{x27}
{\mathcal P}\{\theta \leq t \vert {\mathcal S}_{0}\}\} = W(t \vert
{\mathcal S}_{0}) = {\mathcal P}\{\xi(t) = N_{A} \vert {\mathcal
S}_{0}\} = p(u, N_{A} \vert {\mathcal S}_{0})
\end{equation}
is the probability that the lifetime of the system ${\mathcal S}$
is not larger than $t$ provided that the initial state of the
system was ${\mathcal S}_{0} = \{N_{A}, N_{X}\}$. We can see that
the properties of the system lifetime are determined by the
probability $p(t, N_{A})$.~\footnote{ Here we omitted the notation
referring to the initial state.} The probability density function
of the lifetime is nothing else than
\begin{equation} \label{x28}
w(t \vert {\mathcal S}_{0}) = \frac{dp(t, N_{A})}{dt},
\end{equation}
and we can see that
\begin{equation} \label{x29}
\int_{0}^{\infty} e^{-st}\;w(t \vert {\mathcal S}_{0})\;dt =
\tilde{w}(s \vert {\mathcal S}_{0}) = s\;U_{N_{A}}(s),
\end{equation}
since $p(0, N_{A}) = 0$, if $N_{A} \neq 0$. By using the formula
(\ref{x11}) one obtains
\begin{equation} \label{x30}
\tilde{w}(s \vert {\mathcal S}_{0}) = \frac{h_{0}h_{1} \cdots
h_{N_{A}-1}}{(s+h_{0})(s+h_{1}) \cdots (s+h_{N_{A}-1})},
\end{equation}
and it is evident that
\[ \tilde{w}(0 \vert {\mathcal S}_{0}) =
\int_{0}^{\infty} w(t \vert {\mathcal S}_{0})\;dt = 1. \]

\begin{figure} [ht!]
\protect \centering{
\includegraphics[height=8cm, width=12cm]{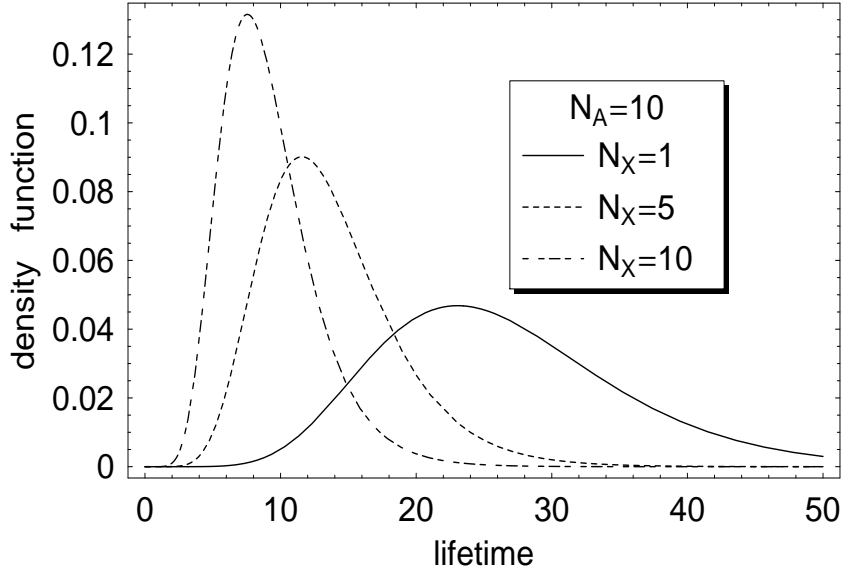}}\protect
\vskip 0.2cm \protect \caption{\label{fig6}{\footnotesize
Probability density function of the lifetime for three initial
states: $\{10, 1\},\;\{10, 5\}$ and $\{10, 10\}$.}}
\end{figure}

It can be worthwhile to study the temporal behavior of systems
containing at $t=0$ only $1$ particle of type $X$ and $N_{A} > 1$
of type $A$, i.e. to follow the process from the initial state
${\mathcal S}_{0} = \{N_{A}, 1\}$ to the dead state ${\mathcal
S}_{d} = \{0, N_{A}+1\}$. Since in this case $N_{A} = n_{c} + 1$
and $N_{X} = 1$, if $n_{c} = 2k_{c}$, where $k_{c}$ is a
nonnegative integer, then we obtain
\[ \tilde{w}(s \vert {\mathcal S}_{0}) = \left[\frac{h_{0}h_{1}
\cdots h_{k_{c}}}{(s+h_{0})(s+h_{1}) \cdots
(s+h_{k_{c}})}\right]^{2}, \] while if $n_{c} = 2k_{c} + 1$, then
we can write
\[ \tilde{w}(s \vert {\mathcal S}_{0}) = \left[\frac{h_{0}h_{1}
\cdots h_{k_{c}}}{(s+h_{0})(s+h_{1}) \cdots
(s+h_{k_{c}})}\right]^{2}\;\frac{h_{k_{c}+1}}{s+h_{k_{c}+1}}. \]

\begin{figure} [ht!]
\protect \centering{
\includegraphics[height=8cm, width=12cm]{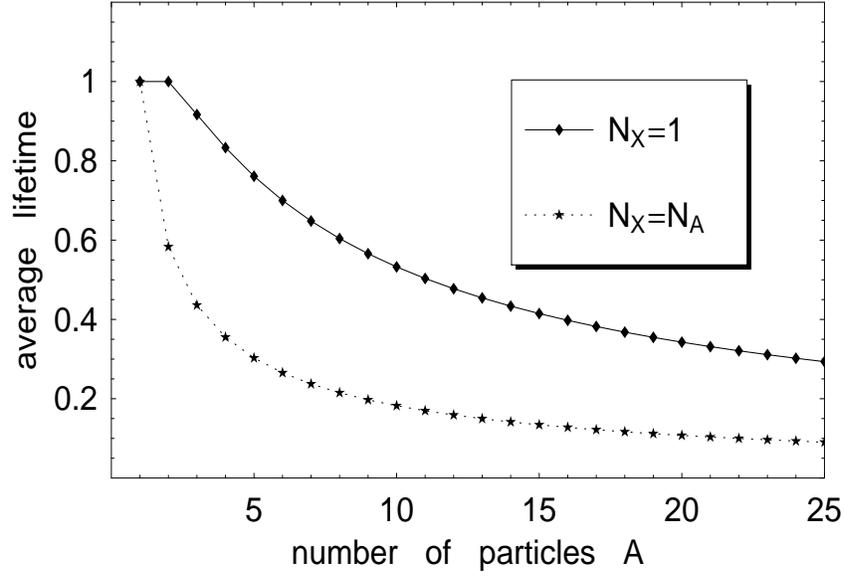}}\protect
\vskip 0.2cm \protect \caption{\label{fig7}{\footnotesize
Dependence of the average lifetime of the system ${\mathcal S}$ on
the initial number of $A$ particles in the cases of $N_{X} = 1$
and $N_{X} = N_{A}$.}}
\end{figure}

\begin{figure} [ht!]
\protect \centering{
\includegraphics[height=8cm, width=12cm]{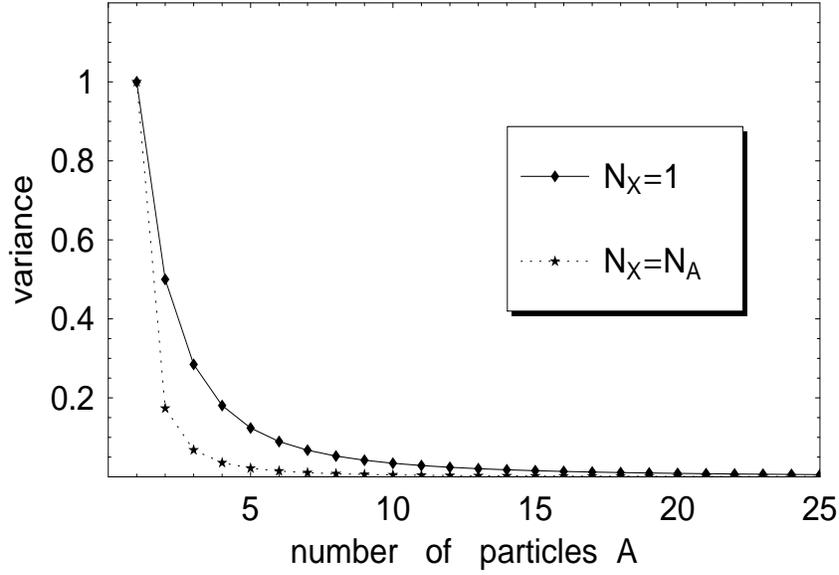}}\protect
\vskip 0.2cm \protect \caption{\label{fig8}{\footnotesize
Dependence of the variance of the lifetime on the initial number
of $A$ particles in the cases of $N_{X} = 1$ and $N_{X} =
N_{A}$.}}
\end{figure}

In order to calculate the probability density function of the
lifetime at different initial states, we should determine the
inverse Laplace transforms of these expressions. In
FIG.~\ref{fig6} we see three probability density functions of the
lifetime $\theta$ belonging to initial states: ${\mathcal S}_{0} =
\{10, 1\},\;\{10, 5\}$ and $\{10, 10\}$. It is remarkable that at
fixed initial number of ${A}$ particles the most probable lifetime
of the system decreases significantly with increasing initial
number of ${X}$ particles.

For the sake of determination of the expectation value ${\bf
E}\{\theta\}$ and the variance ${\bf D}^{2}\{\theta\}$ of the
system lifetime $\theta$ define the function
\begin{equation} \label{x31}
K(s \vert {\mathcal S}_{0}) = \log\;\tilde{w}(s \vert {\mathcal
S}_{0}) = \sum_{n=0}^{N_{A}-1} \log\;\frac{h_{n}}{s+h_{n}}.
\end{equation}
We obtain immediately that
\begin{equation} \label{x32}
{\bf E}\{\theta \vert N_{A}, N_{X}\} = - \left[\frac{dK(s \vert
N_{A}, N_{X})}{ds}\right]_{s=0} = \sum_{n=0}^{N_{A}-1}
\frac{1}{h_{n}},
\end{equation}
and
\begin{equation} \label{x33}
{\bf D}^{2}\{\theta \vert N_{A}, N_{X}\} = \left[\frac{d^{2}K(s
\vert N_{A}, N_{X})}{ds^{2}}\right]_{s=0} = \sum_{n=0}^{N_{A}-1}
\frac{1}{h_{n}^{2}}.
\end{equation}
By using these formulas we calculated the ratios
\[ \frac{{\bf E}\{\theta \vert N_{A}, N_{X}\}}
{{\bf E}\{\theta \vert 1, 1\}}, \;\;\;\;\;\; \text{and}
\;\;\;\;\;\; \frac{{\bf D}^{2}\{\theta \vert N_{A}, N_{X}\}} {{\bf
D}^{2}\{\theta \vert 1, 1\}}\] and plotted them in FIG.~\ref{fig7}
and FIG.~\ref{fig8}.

At the first moment it is surprising that the increase of the
initial number of ${A}$ particles results in a decrease of the
expectation value and the variance of the system lifetime. It
seems to be true that could be happened in the folktale: the
"large" disappears faster than the "small". In fact, the larger is
the initial number of ${A}$ particles the faster is the
autocatalytic reaction, and this explains that we see in figures.

\section{Conclusions}

The main purpose of this paper was to study the stochastic
properties of small systems controlled by autocatalytic reaction.
In order to give exact solution of the problem we assumed the
distribution of reacting particles in the system volume to be
uniform and introduced the notion of \textit{the point model of
reaction kinetics}. In this model the probability of a reaction
between two particles per unit time is evidently proportional to
the product of their actual numbers. For the sake of simplicity we
used in this paper the notations $X$ and $A$ for \textit{the
autocatalytic and the substrate particles}, respectively.

In order to make the calculations not very complex we have chosen
the simplest autocatalytic reaction: $A + X \rightarrow 2 X$, and
constructed a stochastic models for it. The \textit{state of the
system} at a given time moment is completely determined by the
actual numbers of $X$ and $ A$ particles. The system is called
\textit{living} at time $t \geq 0$ when the probability of the
autocatalytic reaction is larger than zero.

We \textit{calculated exactly} the probability $p(t, n)$ of
finding $n$ new ${X}$ particles in the system at time moment $t$,
provided that at $t=0$ the number of new $X$ particles was zero,
and the system contained $N_{X} > 0$ particles of type $X$ and
$N_{A} > 0$ particles of type $A$. We have shown that the
stochastic model results in an equation for the expectation value
$m_{1}(t)$ of the new $X$ particles which differs strongly from
the kinetic rate equation. Consequently, we can state that in this
case \textit{the stochastic model does not support the kinetic law
of the mass action}. It is to mention here that this statement was
already published by R\'enyi \cite{renyi54} many years ago. We
calculated the difference between the time dependencies of mean
values corresponding to the stochastic and the deterministic
models.

\textit{Moment-closure approximations} of two types have been
analyzed, and the results of calculations were compared with the
exact mean values obtained directly from the probabilities
$p_{n}(t),\; n=0,1,\ldots\;$. We found that the "normal"
approximation which neglects the third and higher order cumulants
can be accepted as relatively "good" approximation.

The probability density function of \textit{the system lifetime}
has been also calculated, and it is found that the most probable
lifetime depends sensitively on the number of $X$ particles being
present in the system at time moment $t=0$.

The main conclusion is that \textit{the point model of the
stochastic reaction kinetics} resulted in a deeper understanding
of the random behavior of small systems governed by autocatalytic
processes.

\end{document}